\documentclass[prl,twocolumn,showpacs,superscriptaddress, amsmath,amssymb]{revtex4}

\usepackage{epsfig}
\usepackage{graphicx}
\usepackage{color}
\usepackage{soul}

\begin{document}
\title{Exceptionally strong magnetism in 4$d$ perovskites $R$TcO$_3$ ($R$=Ca, Sr, Ba)}

\author{C. Franchini}
\email[Corresponding author: ]{cesare.franchini@univie.ac.at}
\affiliation{University of Vienna, Faculty of Physics and Center for Computational Materials Science, Vienna, Austria}
\affiliation{Shenyang National Laboratory for Materials Science, Chinese Academy of Science, Shenyang, China}
\author{T. Archer}
\affiliation{School of Physics and CRANN, Trinity College, Dublin 2, Ireland}
\author{Jiangang He}
\affiliation{University of Vienna, Faculty of Physics and Center for Computational Materials Science, Vienna, Austria}
\author{Xing-Qiu Chen}
\affiliation{Shenyang National Laboratory for Materials Science, Chinese Academy of Science, Shenyang, China}
\author{A. Filippetti}
\affiliation{CNR-IOM, UOS Cagliari, Dipartimento di Fisica, Universit\'a di Cagliari, Monserrato (CA), Italy}
\author{S. Sanvito}
\affiliation{School of Physics and CRANN, Trinity College, Dublin 2, Ireland}

\date{\today}

\begin{abstract}
The evolution of the magnetic ordering temperature of the 4$d^3$ perovskites $R$TcO$_3$ ($R$=Ca, Sr, Ba) and its relation 
with its electronic and structural properties has been studied by means of hybrid density functional theory and Monte 
Carlo simulations. When compared to the most widely studied 3$d$ perovskites the large spatial extent of the $4d$ shells 
and their relatively strong hybridization with oxygen weaken the tendency to form Jahn-Teller like orbital ordering. 
This strengthens the superexchange interaction. The resulting insulating G-type antiferromagnetic ground state is 
characterized by large superexchange coupling constants (26-35 meV) and Ne${\rm\acute{e}}$l temperatures (750-1200 K). 
These monotonically increase as a function of the $R$ ionic radius due to the progressive enhancement of the volume and 
the associated decrease of the cooperative rotation of the TcO$_6$ octahedra.
\end{abstract}

%

\maketitle 

Intermediately located between manganese and rhenium, technetium (5$s^2$4$d^5$) shares with its isovalent 
neighbors the intriguing possibility to form oxides with a complex structural, electronic and magnetic phase diagram. 
However, the rare occurrence of natural Tc (Tc is essentially an artificial product of fission reactions) and the related 
radioactive risks (Tc is the lightest radioactive element, whose most abundant isotope, $^{99}$Tc, 
decays with a half-life of 10$^5$ years), have made investigations of Tc-based oxides very 
sparse~\cite{muller64,rodriguez07,rodriguez08,rodriguez11a,avdeev11,rodriguez11b}. Overcoming these difficulties, 
Avdeev {\em et al.}~\cite{avdeev11} and Rodriguez {\em et al.}~\cite{rodriguez11b} have recently reported the 
successful synthesis of the first ever fabricated Tc-based perovskites, namely CaTcO$_3$ and SrTcO$_3$.
Furthermore they have shown that these compounds display the anomalously high Ne${\rm\acute{e}}$l temperatures 
($T_\mathrm{N}$) of 800~K for CaTcO$_3$ and 1000~K for SrTcO$_3$, by far the highest among materials not 
incorporating 3$d$ transition metals. 
These results are surprising and challenge our understanding of the magnetic interaction in perovskites. In particular
they pose three fundamental questions: i) is the origin of such a large magnetic ordering temperature related to the 
strong Tc 4$d$-O $p$ hybridization ?, ii) what is the role played by the structural degrees of freedom? and iii) is the 
enhancement of $T_\mathrm{N}$ upon Ca$\rightarrow$Sr substitution related to the observed increase in unit cell volume 
and the corresponding modification of the internal atomic positions~\cite{kimura03,yamauchi08}~?

In the present letter, by using a combination of hybrid density functional theory~\cite{becke93} and Monte Carlo (MC) 
simulations~\cite{mc}, we address these issues at the microscopic level through a systematic study of the series 
$R$TcO$_3$ ($R$=Ca, Sr, Ba). Our aim here is twofold.  First we wish to interpret and understand the experimental findings
for CaTcO$_3$ and SrTcO$_3$, which can be only captured by beyond-local density functional theory (DFT), due to the 
incorrect treatment of the residual exchange-correlation effects still present in 4$d$ compounds~\cite{maiti05}. 
Secondly we anticipate the experiments predicting the properties of the technetiates series end member BaTcO$_3$.

We have employed the Heyd, Scuseria and Ernzerhof (HSE)\cite{hse} scheme as implemented in the VASP code 
\cite{bpaw,kpaw,vasp1,vasp2}, using a 4$\times$4$\times$4 k-points set, a cutoff energy of 300 eV, a mixing parameter of
$a$=0.1\cite{mix} and standard structural optimization conditions\cite{franchini09}.
\begin{figure}[b]
\includegraphics[width=1.00\columnwidth]{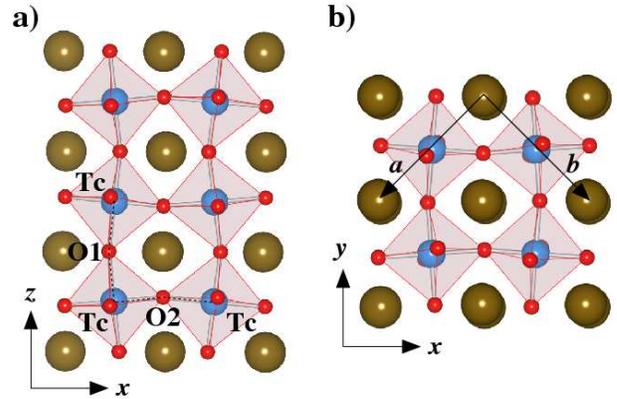}
\caption{(Color online)
Side (a) and top (b) views of the $Pnma$ structure of the distorted perovskite $R$TcO$_3$.
The Tc atoms [medium (light-blue) spheres] are surrounded by O$_6$ octahedran delimited by full (red) lines.
The $R$ atoms are represented by big (dark green) spheres. Both the conventional ($abc$) and cartesian
($xyz$) coordinate systems are shown. In (a) a schematic representation (dashed lines) of the distorted Tc-O
subnetwork indicating the Tc-O1-Tc and Tc-O2-Tc superexchange paths is given, where O1 and O2 are the apical
and planar oxygen atom, respectively.}
\label{fig:fig1}
\end{figure}
$R$TcO$_3$ compounds adopt the same distorted $Pnma$ perovskite structure as LaMnO$_3$~\cite{elemans71}, characterized 
by a GdFeO$_3$-like (GFO) tilting of the TcO$_6$ octahedra caused by the $R$-O and $R$-Tc covalencies, and a 
concomitant Jahn-Teller (JT) Tc-O bond length disproportionation, with long (l) and short (s) Tc-O2 in-plane 
distances and medium (m) Tc-O1 vertical ones (see Fig. \ref{fig:fig1} and Table \ref{tab:1}). When compared to LaMnO$_3$, 
the experimental values of the JT modes $Q_2=2(l-s)/\sqrt(2)$ and $Q_3=2(2m-l-s)/\sqrt(6)$ in CaTCO$_3$ and SrTCO$_3$ 
(see Table \ref{tab:1}) are significantly smaller due to the intrinsically different electronic configuration of the 
transition metal ions. In LaMnO$_3$ the Mn$^{3+}$ ions 3$d$ states are in a $t^3_{2g}e^1_g$ configuration and a strong JT
effect is required to release the degeneracy of the $e_g$ orbitals and to form an orbitally ordered 
state~\cite{pavarini10}. In contrast in $R$TcO$_3$ the Tc$^{4+}$ ions (4$d^3$: $t^3_{2g}e^0_g$) have completely filled 
(empty) $t_{2g}$($e_g$) manifolds, similarly to Mn$^{4+}$ ions in CaMnO$_3$\cite{filippetti02},
a configuration which inhibits the occurence of electronic and structural JT instabilities, and leads to an insulating
G-type antiferromagnetic (AFM-G) ground state\cite{matar03}.
However, at variance with $3d$ transition metal (TM) ions in perovskites, where $d^3$ configurations unavoidably lead 
to fully polarized $m$= 3$\mu_\mathrm{B}$ moments on TM (e.g. CaMnO$_3$), here due to the much larger extension of the 
$4d$ electrons and in turns a much stronger hybridizations with O $p$ states, a sizable amount of charge fills the 
$4d$ minority states as well, with a consequential reduction of magnetic moment and an overtly evident intra-atomic 
Hund's rule violation.

\begin{table}
\caption{
Comparison between the low temperature (3 K)~\cite{lowt}  experimental and calculated structural parameters for the
G-type oriented $R$TcO$_3$ family. The corresponding cooperative Jahn-Teller local modes $Q_2=2(l-s)/\sqrt(2)$ and 
$Q_3=2(2m-l-s)/\sqrt(6)$ are also reported. In the definition $l$, $s$ and $m$ indicate the three inequivalent long (Tc-O2$^l$), 
short (Tc-O2$^s$) and medium (Tc-O1) Tc-O distances, respectively. Room temperature experimental data for CaTcO$_3$ 
and SrTcO$_3$ are from  Refs. \onlinecite{avdeev11} and \onlinecite{rodriguez11b}, respectively. No experimental data are 
available for BaTcO$_3$. As a reference we report here the relevant experimental structural data for the prototypical 
JT/GFO compound LaMnO$_3$: Q$_2$=0.398, Q$_3$=-0.142, $\widehat{{\rm Mn}-{\rm O1}-{\rm Mn}}$=154.3,
$\widehat{{\rm Mn}-{\rm O2}-{\rm Mn}}$=156.7, Mn-O2$^l$=2.184, Mn-O2$^s$=1.903 and Mn-O$_1$=1.957~\cite{elemans71}.} 
\vspace{0.3cm}
\begin{ruledtabular}
\begin{tabular}{lccccc}
   &\multicolumn{2}{c}{CaTcO$_3$}&\multicolumn{2}{c}{SrTcO$_3$}&\multicolumn{1}{c}{BaTcO$_3$}\\
\hline\hline
                                                    & Expt.    & HSE   &Expt.    &HSE    & HSE    \\
                                                    &          &       &         &       &        \\
$a$ (\AA)                                         & 5.526    &5.527  &5.543    &5.559  & 5.678  \\
$b$ (\AA)                                         & 7.695    &7.695  &7.854    &7.856  & 8.038  \\
$c$ (\AA)                                         & 5.389    &5.386  &5.576    &5.592  & 5.688  \\
V (\AA$^3$)                                         & 229.15   &229.04 &242.74   &244.20 & 259.59 \\
$\widehat{{\rm Tc}-{\rm O1}-{\rm Tc}}$ ($^{\circ}$) & 150.43   &150.52 &161.57   &164.74 & 178.89 \\
$\widehat{{\rm Tc}-{\rm O2}-{\rm Tc}}$ ($^{\circ}$) & 151.53   &150.84 &166.96   &168.76 & 179.22 \\
Tc-O2$^l$ (\AA)                                   & 2.002    & 1.998 &2.015    &1.983  & 2.010  \\
Tc-O2$^s$ (\AA)                                   & 1.990    & 1.993 &1.942    &1.979  & 2.010  \\
Tc-O1 (\AA)                                       & 1.985    & 1.988 &1.990    &1.981  & 2.010  \\
Q$_2$                                               & 0.017    & 0.007 &0.103    &0.005  & 0.000  \\
Q$_3$                                               &--0.018   &-0.012 &0.018    &0.001  & 0.000  \\
\end{tabular}
\end{ruledtabular}
\label{tab:1}
\end{table}

The structural parameters obtained by a full geometry optimization for the AFM-G ground 
state (Table \ref{tab:1}) are in excellent agreement with the available low temperature experimental data and 
the overall relative error is smaller than 1\%. The Tc-O2 distances in SrTcO$_3$ are the only exceptions since a deviation
of almost 2\% is observed. Notice that, at variance with CaTcO$_3$, in SrTcO$_3$ experiment finds an appreciable 
planar bond length anisotropy, i.e. the two Tc-O2 bond lengths differ by about 0.07~\AA~(at 3 K)~\cite{rodriguez11b,lowt}. 
This appears quite anomalous if one considers that the 5\% volume expansion associated to the substitution of Ca with the
heavier Sr is expected to induce the quenching of the cooperative $Q_2$ and $Q_3$ JT modes. This translates in a reduction
of the Tc-O disproportionation, similarly to the trend observed in the most widely study 
$R$MnO$_3$ series~\cite{kimura03, yamauchi08, trimarchi05}.
The relatively strong Tc-O2 bond length anisotropy found in the experiments is comparable to that found in LaMnO$_3$, 
and would suggest the presence of an orbitally ordered state in SrTcO$_3$. This is unexpected for a 4$d$ perovskite, 
given the abovementioned extendend character of the 4$d$ manifold. On the other hand, HSE yields a Tc-O2 
bond length anisotropy smaller for SrTcO$_3$ (0.005) than for CaTcO$_3$ (0.007), thus drawing a globally 
consistent picture.
Our results describe clear trends going from small (Ca) to large (Ba) cation size: the increase of volume, the stretching 
of $\widehat{{\rm Tc}-{\rm O}-{\rm Tc}}$ bond angles, the consequential decrease of tilting distortion and the 
quenching of the Q$_2$ mode, which controls the relative difference between the in-plane Tc-O2 bond lengths.    

How these subtle structural properties affect the magnetism and how they can explain the observed 
surprisingly large $T_\mathrm{N}$ is discussed next. 

\begin{table}
\caption{
Compilation of measured and calculated magnetic moment $m$, exchange coupling constants,
T$_N$ and relative stability of the different magnetic phases considered ($\Delta{E}_\mathrm{GC}$=E(AFM-C)-E(AFM-G), 
$\Delta{E}_\mathrm{GA}$=E(AFM-A)-E(AFM-G) and $\Delta{E}_\mathrm{GF}$=E(FM)-E(AFM-G)) for $R$TcO$_3$ 
($R$=Ca, Sr and Ba). Experimental data ($m$, taken at 4 K, and $T_\mathrm{N}$) are only available for 
CaTcO$_3$~\cite{avdeev11} and SrTcO$_3$~\cite{rodriguez11b}.} 
\vspace{0.3cm}
\begin{ruledtabular}
\begin{tabular}{lccccc}
   &\multicolumn{2}{c}{CaTcO$_3$}&\multicolumn{2}{c}{SrTcO$_3$}&\multicolumn{1}{c}{BaTcO$_3$}\\
\hline\hline
                                                     & Expt.      & Calc.  & Expt.  & Calc.  & Calc.      \\
                                                     &            &        &        &        &            \\
$m$ ($\mu_{\rm B}$)                                  & $\sim$ 2.0 & 2.10   & 2.13   & 2.04   & 2.04       \\
$\Delta{E}_\mathrm{CG}$ (meV/f.u.)                   & --         &  81    & --     & 121    & 139        \\
$\Delta{E}_\mathrm{AG}$ (meV/f.u.)                   & --         & 191    & --     & 289    & 299        \\
$\Delta{E}_\mathrm{FG}$ (meV/f.u.)                   & --         & 313    & --     & 449    & 439        \\
$J_1$ (meV)                                          & --         & -26.2  & --     & -35.3  & -34.0      \\
$J_2$ (meV)                                          & --         & -1.3   & --     & -0.7   &  0.1       \\
$T_{\rm N}$ (K)                                      & 800        & 750    & 1020   & 1135   &  1218      \\
\end{tabular}
\end{ruledtabular}
\label{tab:2}
\end{table}

By mapping the HSE total energies for different magnetic configurations onto a Heisenberg Hamiltonian (for unitary spins) 
we have evaluated the nearest-neighbor (NN, $J_1$) and next-nearest-neighbor (NNN, $J_2$) magnetic exchange parameters as 
a function of the ionic radius $r_R$\cite{xqc09}. The results are summarized in Fig.~\ref{tab:2}.
The NN superexchange (SE) Tc-O-Tc path, as illustrated in Fig.~\ref{fig:fig1}, connects 
NN Tc atoms via planar (Tc-O2-Tc) and apical (Tc-O1-Tc) oxygen atoms. Because of the almost identical Tc-O distances and 
$\widehat{{\rm Tc}-{\rm O}-{\rm Tc}}$ angles  (see Table~\ref{tab:1}) we have found that the in-plane ($J^{\parallel}_1$) 
and out-of-plane ($J^{\perp}_1$) values of $J_1$ are equal, both in sign (thus establishing the AFM-G ordering) and 
magnitude (within 10$^{-1}$ meV). Therefore we have finally adopted an effective Hamiltonian including an isotropic $J_1$
magnetic interaction.

In addition to the ferromagnetic (FM) alignment three additional AFM configurations have been considered in fitting the
$J$'s: AFM-A [(100) planes of like spins alternating along the $c$ axis], AFM-C (antiferromagnetic order in the $xy$ plane 
with a ferromagnetic order along $z$) and AFM-G (all spins are antiferromagnetically coupled to their nearest neighbors). 
At variance with what found with conventional local DFT functionals~\cite{avdeev11,rodriguez11b}, HSE provides very stable
magnetic solutions for every spin ordering with almost identical values of the magnetic moment of Tc 
($\approx$ 2 $\mu_B$, see Table \ref{tab:2}). 
Furthermore the AFM-G phase turns out to be the most stable (see Table \ref{tab:2}). 
The Heisenberg Hamiltonian is then used to evaluate $T_\mathrm{N}$ via MC 
calculations~\cite{fisher09}. Note that in doing that we have rescaled the $J$'s by a factor $S^2/S(S+1)$ (S=3/2), which 
accounts for quantum fluctuations\cite{wan06}. The results are summarized in Table \ref{tab:2}. The calculated 
$T_\mathrm{N}$ for CaTcO$_3$ (750~K) and SrTcO$_3$ (1135~K) are in very good agreement with experiments and the 
$T_\mathrm{N}$ predicted for BaTcO$_3$ (1218~K), if experimentally confirmed, would represent the highest 
magnetic ordering temperature for any system without 3$d$ states. 

\begin{figure}
\includegraphics[clip,width=0.9\columnwidth]{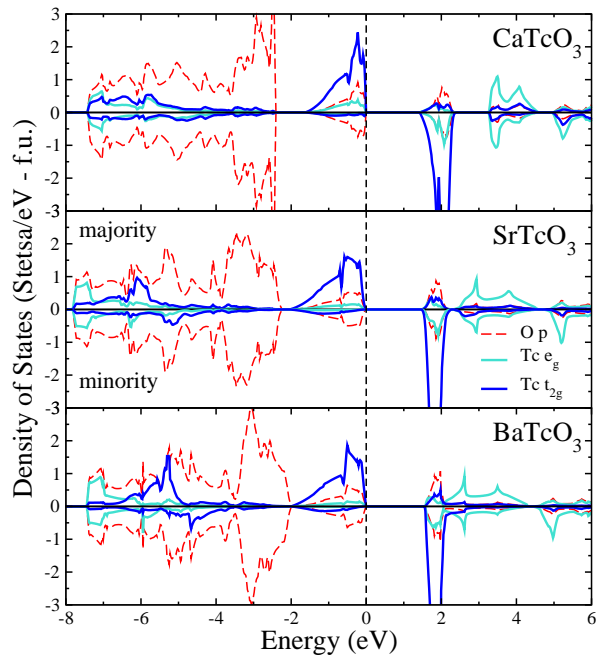}          
\caption{(Color online) HSE calculated density of states for the $R$TcO$_3$ series in the G-type AFM ground state
decomposed over majority and minority O $p$ and Tc t$_{2g}$ and e$_{g}$ states. The calculated band gaps are: 
1.4~eV (CaTcO$_3$), 1.48~eV (SrTcO$_3$) and 1.57~eV (BaTcO$_3$).}   
\label{fig:fig2}
\end{figure}

These unexpectedly large magnetic ordering temperatures can be understood in the framework of Anderson's 
theory of SE interactions~\cite{anderson59}, which links the strength of the SE coupling constant with the actual
hybridization between the metal and the mediating atom, and Van Vleck's theory of antiferromagnetism~\cite{vanvleck41}, 
which connects the strength of the SE interaction with the magnetic ordering temperature. The calculated inter-atomic 
NN and NNN coupling constants listed in Table~\ref{tab:2} show that the antiferromagnetic NN $J_1$ is the dominating 
parameter ($\sim$~-30 meV) and that it is almost two orders of magnitude larger than $J_2$ ($\sim$ -0.5 meV). 
The calculated density of states (DOS) displayed in Fig.~\ref{fig:fig2} shows that these huge $J_1$ value arises 
from the strong covalency between the Tc $t_{2g}$ and O $p$ orbitals evolving along the wide 4$d$ $t_{2g}$ manifold, 
in particular for the topmost valence states spreading from -1.5/2~eV to the Fermi level ($E_\mathrm{F}$). 
The increasing bandwidth ($w$) of this group of hybridized bands observed when going from CaTcO$_3$ ($w=1.5$ eV) 
to SrTcO$_3$ and BaTcO$_3$ ($w=2.0$ eV) associated with the enhanced $t_{2g}$-$p$ hybridization in the 3~eV 
wide $t_{2g}$ band around -5.5~eV explains the larger $J_1$ and the corresponding larger $T_\mathrm{N}$ for 
SrTcO$_3$ and BaTcO$_3$. At this point a fundamental question naturally arises: {\em Why the Tc-O hybridization 
increases along the $R$TcO$_3$ series when $r_R$ get larger?}

\begin{figure}
\includegraphics[clip, width=0.9\columnwidth]{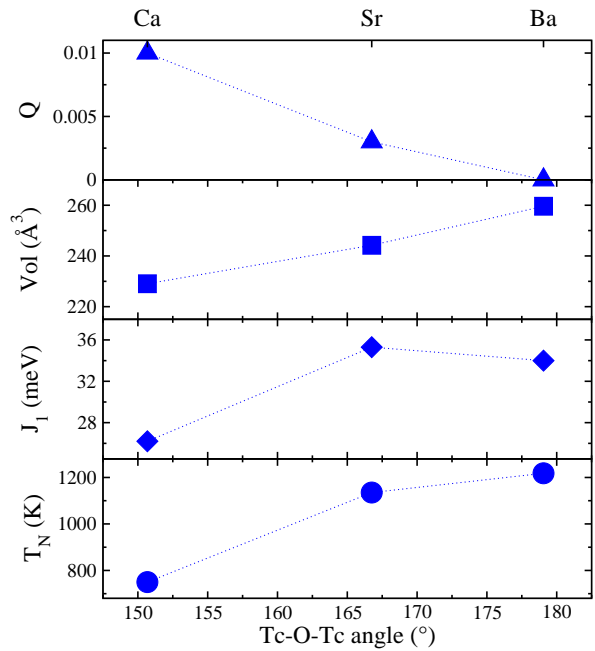}
\caption{(Color online) Dependency of the relevant magnetic ($T_\mathrm{N}$ and $J_1$) and structural [Volume
and Q=$1/2$(Q$_1$+Q$_2$)] quantities on the $\widehat{{\rm Tc}-{\rm O}-{\rm Tc}}$ angle (average between
$\widehat{{\rm Tc}-{\rm O1}-{\rm Tc}}$ and $\widehat{{\rm Tc}-{\rm O2}-{\rm Tc}}$) in the
$R$TcO$_3$ series.
}
\label{fig:fig3}
\end{figure}

In order to answer to this question we analyse the coupling between the structural changes, caused by the substitution of 
Ca with bigger isovalent atoms (Sr and Ba), and the electronic and magnetic properties, as exemplified in 
Fig.~\ref{fig:fig3}. We recall, that the most significant effects on the crystal structure caused by increasing $r_R$ are 
(i) the volume enhancement, (ii) the quenching of the JT distortions Q$_2$ and Q$_3$ and (iii) the decrease               
of the cooperative rotation of the TcO$_6$ octahedra represented by the $\widehat{{\rm Tc}-{\rm O}-{\rm Tc}}$ bond 
angles. The angles $\widehat{{\rm Tc}-{\rm O1}-{\rm Tc}}$ and $\widehat{{\rm Tc}-{\rm O2}-{\rm Tc}}$ are crucial 
quantities to explain the evolution of the SE interactions and of the magnetic ordering temperature. In fact, the 
monotonic increase of the $\widehat{{\rm Tc}-{\rm O}-{\rm Tc}}$ bond angles (in brief the average between $\widehat{{\rm Tc}-{\rm O1}-{\rm Tc}}$ and 
$\widehat{{\rm Tc}-{\rm O2}-{\rm Tc}}$, see Fig.~\ref{fig:fig3}) leads to the progressive rectification of the NN 
superexchange paths. This generates, in a tight-binding framework, an enhanced Tc-$t_{2g}$/O-$p$ hybridization, as 
confirmed by the DOS (see Fig.~\ref{fig:fig2}). As displayed in Fig.~\ref{fig:fig3}, $T_\mathrm{N}$ steeply increases 
from CaTcO$_3$ (750~K) to SrTcO$_3$ (1135~K) as a consequence of the observed larger change in 
$\widehat{{\rm Tc}-{\rm O}-{\rm Tc}}$, which goes from 151$^{\circ}$ to 167$^{\circ}$. When moving from SrTcO$_3$ to 
BaTcO$_3$ (1218~K) the rise of $T_\mathrm{N}$ is weaker due to a smaller change of $\widehat{{\rm Tc}-{\rm O}-{\rm Tc}}$ 
(from 167$^{\circ}$ to 179$^{\circ}$) and to a further reduction and sign change in $J_2$. 
In analogy with the RMnO3 perovskites, the increase of $T_\mathrm{N}$ for larger
Tc-O-Tc angles correlates with a progressive reduction of the JT dis-
tortions (i.e. a decrease of the associated structural ordering tempera-
ture).\cite{kimura03}

In conclusion we remark that our findings for the 4$d$ $R$TcO$_3$ series are consistent with the Goodenough-Kanamori 
rules~\cite{GoodenoughKanamori} and follow the general trend observed in 3$d$ $R$MnO$_3$, namely that $T_\mathrm{N}$ 
increases by increasing the $\widehat{{\rm Tc}-{\rm O}-{\rm Tc}}$ angles.
The remarkably different $T_\mathrm{N}$'s in manganites ($T_\mathrm{N}<$~150~K) and technetiates ($T_\mathrm{N}>$~750~K) 
can be explained with the reduced spatial extension of 3$d$ shell, which suppresses the $d$-$p$ hybridization in manganites 
and thereby inhibits strong SE interactions. We also emphasize, that the correct description of this complex class of
materials can be only achieved within a beyond-local functional method such as HSE. 
HSE is capable of capturing the delicate balance 
between the SE mechanism and both, the Hund's coupling (unlike local functionals, HSE provides well defined magnetic 
solutions for different spin configurations) and the intra-atomic Coulomb repulsion (though substantially reduced with 
respect to 3$d$ perovskites,  on site Coulomb interaction survives in 4$d$ technetiates and contributes to the formation 
of a rather large gap which local functionals seriously underestimates).

\section*{Acknowledgements}

This work was supported by the EU-FP7 within the EU-INDIA project ATHENA. Supercomputer time was provided by the Vienna 
Scientific Cluster. C. F. acknowledges grant support from the CAS (Fellowship for Young International Scientists) and the NSFC 
(Grand Number: 51050110444). We are grateful to B.-J. Kennedy for sharing with us the unpublished low temperature structural 
data.


\begin{thebibliography}{}
\bibitem{mix}
The value of the mixing parameter $a$=0.1 is reduced with respect
to the standard value (0.25) since self-interaction error
are considered to be smaller in 4$d$ transition metal oxides.
Se Ref. \onlinecite{maiti05}
\end{thebibliography}
\end{document}